\documentclass[aps,amsmath,amssymb,groupedaddress,twocolumn,showpacs]{revtex4}
\usepackage{graphicx}
\usepackage{epsfig}

\newcommand{\nc}{\newcommand}
\nc{\beq}{\begin{equation}}
\nc{\eeq}{\end{equation}}
\nc{\beqa}{\begin{eqnarray}}
\nc{\eeqa}{\end{eqnarray}}

\def\gsim{\mathrel{\rlap{\lower4pt\hbox{\hskip1pt$\sim$}}
    \raise1pt\hbox{$>$}}}       %greater than or approx. symbol

\begin{document}

\title{Quantum gravity at a TeV and the renormalization of Newton's constant}

\author{Xavier Calmet$^{a,b}$\footnote{Charg\'e de recherches du F.R.S.-FNRS.}}\email{calmet@uoregon.edu}
\author{Stephen~D.~H.~Hsu$^{a}$} \email{hsu@uoregon.edu}
\author{David Reeb$^{a}$,} \email{dreeb@uoregon.edu}
\affiliation{$^{a}$Institute of Theoretical Science, University of Oregon,
Eugene, OR 97403, USA \\
$^{b}$Catholic University of Louvain,
Center for Particle Physics and Phenomenology,
2, Chemin du Cyclotron,
B-1348 Louvain-la-Neuve, Belgium}

\begin{abstract}
We examine whether renormalization effects can cause Newton's constant
to change dramatically with energy, perhaps even reducing the scale of
quantum gravity to the TeV region without the introduction of extra
dimensions. We examine a model that realizes this possibility and describe experimental signatures from the production of small black holes.
\end{abstract}

\date{June 2008}


\maketitle

\date{today}
\bigskip

It has become conventional to interpret the Planck scale $M_P$ as a fundamental scale of Nature, indeed as the scale at which quantum gravitational effects become important. However, Newton's constant $G$ ($G = M_P^{-2}$ in natural units~$\hbar = c = 1$) is measured in very low-energy experiments, and its connection to physics at short distances -- in particular, quantum gravity -- is tenuous, as we explore in this paper. 

\begin{figure}[tb]
\center
\includegraphics[scale=0.5]{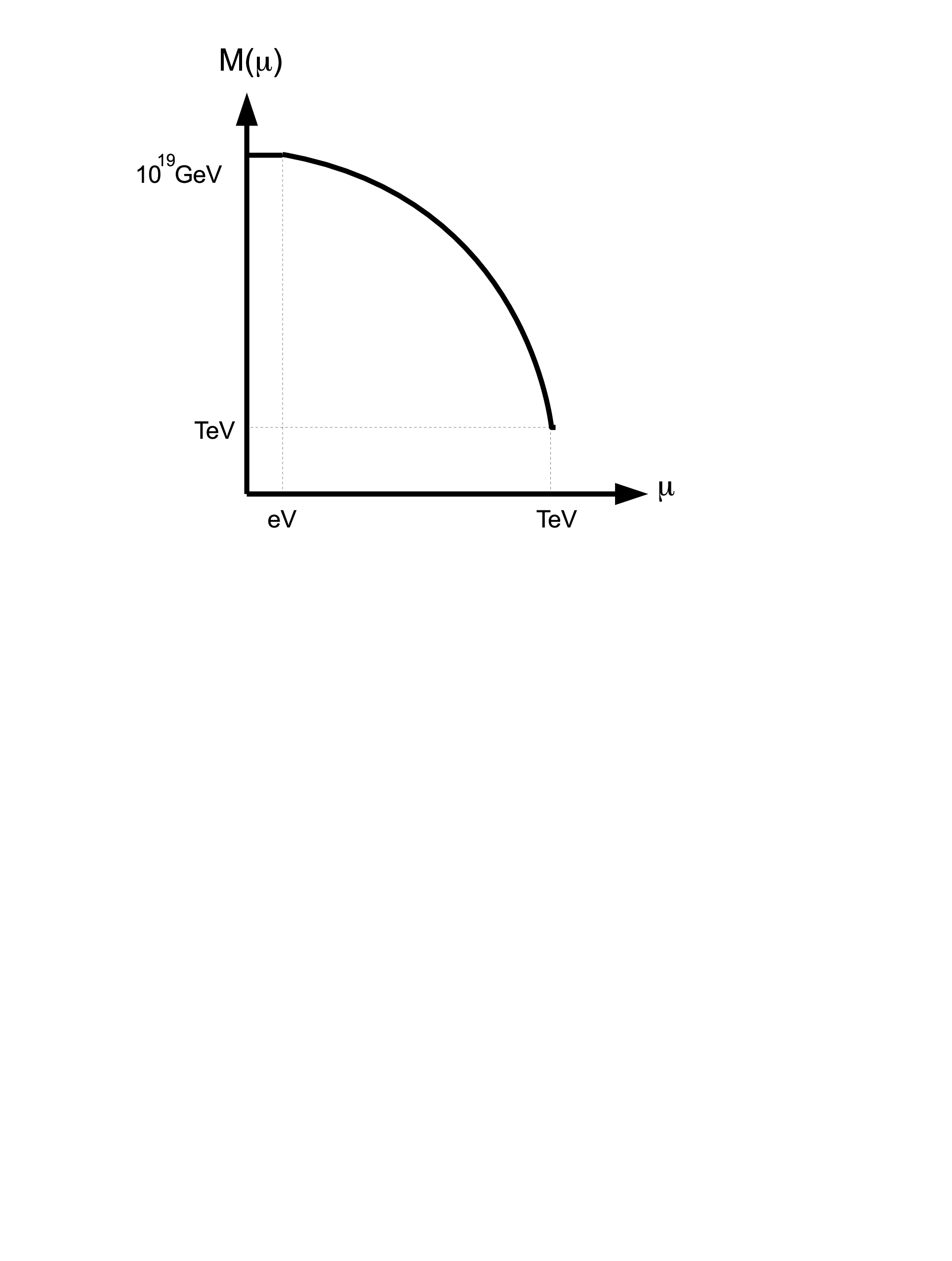}
\caption{Schematic illustration of a possible renormalization group evolution of $M$ with the scale $\mu$.}
\label{fig:test}
\end{figure}

If the strength of gravitational interactions (henceforth, $G ( \mu )$) is scale dependent, the true scale $\mu_*$ at which quantum gravity effects are large is one at which
\begin{equation}
\label{strong}
G (\mu_*) \sim \mu_*^{-2}~.
\end{equation} 
This condition implies that fluctuations in spacetime geometry at length scales $\mu_*^{-1}$ will be unsuppressed. Below we will show that (\ref{strong}) can be satisfied in models with $\mu_*$ as small as a TeV (see Fig.~\ref{fig:test}). Gravity has only been tested at distances greater than that corresponding to an energy scale of $10^{-3}\,$eV. New physics in the form of particles with masses greater than this scale or of modifications to gravity itself could lead to this running of Newton's constant. In such models there is no hierarchy problem, and quantum gravity can be probed by experiments at TeV energies. It is well known that this can be the case in extra-dimensional models \cite{ArkaniHamed:1998rs}, but is this also possible in four dimensions?

Note, we will sometimes refer to an effective Planck scale $M( \mu )$ defined by $G( \mu ) = M ( \mu )^{-2}$. Then, the quantum gravity condition (\ref{strong}) is simply $M( \mu_* ) \sim \mu_*$.

\begin{figure}
%[htp]
\centering
\includegraphics[width=3in]{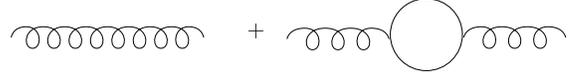}
\caption{Contributions to the running of Newton's constant.}\label{Figure2}
\end{figure}

We will now give a heuristic description of how significant scale dependence of $G$ can arise. A more technical derivation will be given later in the paper. The basic ingredients, screening due to quantum fluctuations and renormalization group evolution, are familiar from QCD. We consider one scalar field coupled to gravity and adopt the following notation:
\begin{equation}
\label{action}
S=\int d^4x \sqrt{-g} \left (-\frac{1}{16\pi G}R + \frac{1}{2} g^{\mu\nu}\partial_\mu \phi \partial_\nu \phi  \right)~.
\end{equation}
Consider the gravitational potential between two heavy, non-relativistic sources, which arises through graviton exchange (Fig.~\ref{Figure2}). 
The leading term in the gravitational Lagrangian is 
$G^{-1} R \sim G^{-1} h \Box h$ with $g_{\mu \nu} = \eta_{\mu \nu} + h_{\mu \nu}$. By not absorbing $G$ into the definition of the small fluctuations $h$ we can interpret quantum corrections to the graviton propagator from the loop in Fig.~\ref{Figure2} as a renormalization of $G$. Neglecting the index structure, the graviton propagator with one-loop correction is
\begin{equation}
\label{oneloopcorrection}
D_h (q) ~\sim~ \frac{i\,G}{q^2} ~+~ \frac{i\,G}{q^2} \Sigma \frac{i\,G}{q^2} ~+~ \cdots ~ ,
\end{equation}
where $q$ is the momentum carried by the graviton. The term in $\Sigma$ proportional to $q^2$ can be interpreted as a renormalization of $G$, and is easily estimated from the Feynman diagram:
\begin{equation}
\label{loop}
\Sigma ~\sim~ -i q^2 \int^\Lambda d^4p~ D(p)^2 p^2 ~+~ \cdots ~ ,
\end{equation}
where $D(p)$ is the propagator of the particle in the loop. In the case of a scalar field the loop integral is quadratically divergent, and by absorbing this piece into a redefinition of G in the usual way one obtains an equation of the form
\begin{equation}
\label{RG1}
\frac{1}{G_{\rm ren}} ~=~ \frac{1}{G_{\rm bare}} + c \Lambda^2~,
\end{equation}
where $\Lambda$ is the ultraviolet cutoff of the loop and $c \sim 1/16 \pi^2$. $G_{\rm ren}$ is the renormalized Newton constant measured in low-energy experiments. Fermions contribute with the same sign to the running of Newton's constant, whereas gauge bosons contribute with the opposite sign than scalars (see below).

Taking $\Lambda = \mu_*$ (so that the loop cutoff coincides with the onset of quantum gravity) gives $G_{\rm bare} = G( \Lambda ) = \mu_*^{-2}$, and then demanding $G_{\rm ren} = M_P^{-2}$ implies that $\mu_*$ cannot be very different from the Planck scale $M_P$ unless $c$ is very large. For example, to have $\mu_*  \sim {\rm TeV}$ requires $c \sim 10^{32}$: it takes  $10^{32}$ ordinary scalars or fermions with masses below 1 TeV (which can run in the loop) to produce the evolution in Fig.~\ref{fig:test}. This observation has already been made by Dvali {\it et al.}~\cite{Dvali:2001gx,Dvali:2007hz,Dvali:2007wp}, although in \cite{Dvali:2007hz} the argument is expressed in terms of a consistency condition from black hole evaporation rather than as renormalization group behavior. Their argument is as follows. Consider a model with $N$ different types of $Z_2$ charges, each of which is the remnant of a gauge symmetry, so that each has long range quantum hair. Assume the $Z_2$ charge carriers all have mass $m$, and let a black hole form from $N$ (one of each) of these particles. The black hole cannot radiate (very much) $Z_2$ charge until: $T \sim M_P^{\,2}/ M_{bh}  \sim m$, where $M_{bh}$ is the mass of the hole. To radiate all $N$ units of $Z_2$ charges, it is necessary that $M_{bh} \sim M_P^{\,2} / m > N m$, which implies $M_P^{\,2} > N m^2$.  
Note that this argument makes two nontrivial assumptions about quantum gravity: unitarity of black hole evaporation and the absence of black hole remnants.  The renormalization calculation, requiring fewer assumptions, shows that additional spin 0 or 1/2 particles of any mass less than $\mu_*$  tend to weaken gravity in the infrared. The hierarchy problem can be solved without implying the relation $m \sim M_P / \sqrt{N}$.

If the effective Planck scale evolves with energy scale, one might ask which is the relevant $M_P$ for Dvali {\it et al.}'s consistency condition? Roughly speaking, it is the Planck scale evaluated at the length scale of the Schwarzschild radius: $R = M_{bh} / M_P^{\,2}$. In our results, most of the RG evolution happens near the scale $\mu_*$ and $M_P$ rapidly reaches its ultimate low energy value of $10^{19}$ GeV. As long as the black hole is larger than $\mu_*^{-1}$ (as is necessary for a semi-classical description), the two pictures are consistent.

As mentioned, the new fields contributing to $c$ in (\ref{RG1}) can have masses as small as $\sim 10^{-3}$ eV. However,  there is a large hierarchy between $10^{-3}$ eV and 1 TeV, and the mass of such a light scalar would not be stable. One could invoke supersymmetry in order to stabilize the masses of the $N$ light spin-$0$ particles. As mentioned above, spin-$1/2$ and spin-$0$ particles contribute to the running of Newton's constant with the same sign. 

The number of new degrees of freedom we are required to introduce may seem outrageous, but it is of the same order as in models with large extra-dimensions \cite{ArkaniHamed:1998rs}. In such models the higher-dimensional action is of the form
\begin{equation}
S=\int d^4x \, d^{d-4}x' \, \sqrt{-g} ~ \left( M_*^{d-2} \, {\cal R} ~+~ \cdots ~~ \right)~,
\end{equation}
so that the effective $3+1$ dimensional Planck scale is given by  $M_P^{\,2} = M_*^{d-2} V_{d-4}$, where $V_{d-4}$ is the volume of the extra dimensions and $M_*$ is the $d$-dimensional Planck mass. By taking $V_{d-4}$ large, $M_P$ can be made of order $10^{19}$ GeV while $M_*\sim$ TeV, however the number of additional degrees of freedom in the bulk is of order $V_{d-4} M_*^{d-4} \sim 10^{32}$.

%%%%%%%%%%%%%%%%%%%%%%%%%%%%%%%%%%%%%%%
% "The rest we leave to the model freaks!" --- M. Veltman
%%%%%%%%%%%%%%%%%%%%%%%%%%%%%%%%%%%%%%%

Now we will give a functional derivation of equation (\ref{RG1}), which shows that the sign of the contribution of the scalar fields to the running of Newton's constant is not an artifact of the crude (non-covariant) regularization procedure we used earlier. Consider the contribution of a scalar field minimally coupled to gravity. We follow the presentation of Larsen and Wilczek \cite{LarsenWilczek} (see also \cite{DeWitt,BirrellDavies}). 
The one-loop effective action $W$ is defined through
\begin{eqnarray}
e^{-W}&=&\int {\cal D}\phi~e^{-\frac{1}{8\pi}\int\phi (-\Delta+m^2)\phi} \\ 
\nonumber
&=&[{\rm det}(-\Delta+m^2)]^{-\frac{1}{2}}~.
\end{eqnarray}
We define the heat kernel
\begin{eqnarray}
H(\tau) \equiv {\rm Tr}e^{-\tau\Lambda}=\sum_i e^{-\tau\lambda_i}~,
\end{eqnarray}
where $\lambda_i$ are the eigenvalues of $\Lambda=-\Delta + m^2$.
Then the effective action reads
\begin{eqnarray}
\label{W}
W=\frac{1}{2}{\rm ln}\,{\rm det}\Lambda=\frac{1}{2}\sum_i {\rm ln}\lambda_i
=-\frac{1}{2}\int_{\epsilon^2}^\infty
d\tau \frac{H(\tau)}{\tau}~.
\end{eqnarray}
The integral over $\tau$ is divergent and has to be regulated by an ultraviolet cutoff $\epsilon^2$.
The heat kernel method can be used to regularize the leading divergence of this integral. This technique does not violate general coordinate invariance. One can write
\begin{eqnarray}
H(\tau)=\int dx~G(x,x,\tau)~,
\end{eqnarray}
where
the Green's function $G(x,x^{\prime},\tau)$ satisfies the differential equation
\begin{eqnarray}
(\frac{\partial}{\partial\tau}-\Delta_x)G(x,x^{\prime},\tau)&=&0~; \\
G(x,x^{\prime},0)&=&\delta(x-x^\prime)~.
\end{eqnarray}
In flat space one has
\begin{eqnarray}
G_0(x,x^\prime,\tau) = \left(\frac{1}{4\pi\tau}\right)^{2}
\exp{\left (-\frac{1}{4\tau}(x-x^\prime)^2 \right )}~,
\end{eqnarray}
but in general one must express the covariant Laplacian in local coordinates
and expand for small curvatures.

The result is \cite{BalianBloch}
\begin{eqnarray} \label{BB}
H(\tau)&=& \frac{1}{(4\pi\tau)^2}    
\Big( \int d^4x\, \sqrt{-g}
\\ \nonumber &&~+~ \frac{\tau}{6} \int d^4x \, \sqrt{-g} \, R   
~+~ {\cal O}(\tau^\frac{3}{ 2})\Big)~.
\end{eqnarray}
Plugging this back into (\ref{W}) and comparing to (\ref{action}), one obtains the renormalized Newton constant
\begin{eqnarray} \label{cutdependence}
\frac{1}{G_{\rm ren}} ~=~
 \frac{1}{G_{\rm bare} } + \frac{1}{12 \pi \epsilon^2}~,
\end{eqnarray}
so that $G_{\rm ren}$, relevant for long-distance measurements, is much smaller than the bare value if the scalar field is integrated out ($\epsilon \rightarrow 0$).

Up to this point our results have been in terms of old-fashioned renormalization: we give a relation between the physical observable $G_{\rm ren}$ and the bare coupling $G_{\rm bare}$. A modern Wilsonian effective theory would describe modes with momenta $\vert k \vert < \mu$. Modes with $\vert k \vert > \mu$ have been integrated out and their virtual effects already absorbed in effective couplings $g( \mu )$. In this language, $G_{\rm ren} = G ( \mu = 0 )$ is appropriate for astrophysical and other long-distance measurements of the strength of gravity.

A Wilsonian Newton constant $G( \mu )$ can be calculated via a modified version of the previous method, this time with an infrared cutoff $\mu$. For example, (\ref{W}) is modified to
\begin{eqnarray}
\label{W2}
W=-\frac{1}{2}\int_{\epsilon^2}^{\mu^{-2}}
d\tau \frac{H(\tau)}{\tau}~.
\end{eqnarray}
The resulting Wilsonian running of Newton's constant is
\begin{eqnarray}
\frac{1}{G ( \mu )} ~=~
\frac{1}{G(0) } - \frac{\mu^2}{12 \pi}~,
\end{eqnarray}
 or
 \begin{eqnarray}
 \label{Nrunning}
\frac{1}{G ( \mu )} ~=~
\frac{1}{G(0) } - N \frac{\mu^2}{12 \pi} 
\end{eqnarray}
for $N$ scalars or Weyl fermions, as can be shown by a similar functional calculation. Compare with Larsen and Wilczek in \cite{LarsenWilczek}, who also derive the opposite sign in the gauge boson case.

We note that (\ref{cutdependence}) and (\ref{Nrunning}) are only valid to leading order in perturbation theory. As we near the scale of strong quantum gravity $\mu_*$ we lose control of the model. However, it seems implausible that the sign of the beta function for Newton's constant will reverse, so the qualitative prediction of weaker gravity at low energies should still hold.

There are other quantum corrections from the new particles: the cosmological constant is renormalized as well, as can be seen from equation (\ref{BB}). The relation is of the form
\begin{eqnarray}
\Lambda_{\rm ren} = \Lambda_{\rm bare}+(N_b -N_f) \frac{c'}{\epsilon^4}~,
\end{eqnarray}
where here fermions and bosons contribute oppositely.
The natural value of $|\Lambda_{\rm bare}|$ is of the order of a ${\rm TeV}^4$ since this is the cutoff  we impose on the model, whereas the observed cosmological constant $\sim (10^{-3} \ \mbox{eV})^4$ is much smaller. The $N$ degrees of freedom thus make the problem much more severe, unless we assume the number of new bosons to be nearly equal to that of new fermions. This leads to the intriguing possibility that the hidden sector could be a simple Wess-Zumino model.

The $N$ new degrees of freedom are assumed to be singlets and to couple to the standard model only gravitationally. Graviton loops will typically lead to operators of the type $\phi_i \phi_i  \phi_j \phi_j m_i^2 m_j^2/M(m)^4$ times some logarithmic divergence, where $m$ is the mass of the scalars. If the mass of the scalar field is much smaller than the Planck scale, these operators are strongly suppressed. 
If we choose $m_i\sim 1\,{\mbox{TeV}}$, the factor $m_i^2 m_j^2/M(m)^4$ could naively be of order one, however one has to keep in mind that the running of Newton's constant happens only between $m$ and $\mu_*$ and thus very fast. So we can choose $m$ just smaller than $\mu_*$ and discard these operators.

It seems possible that the large number of hidden degrees of freedom we are introducing could be mimicked, insofar as their effect on the renormalization group equations, by a modification of general relativity of the type $\int d^4x \sqrt{-g} f(R)$, where $f(R)$ is a function of the Ricci scalar: $f(R) = - c_1 R + c_2 R^2 + \cdots~$. For example, if the 
$N$ new particles are all heavy, with $m \sim \mu_*$, then integrating them out would lead to an effective Lagrangian of this type at scales $\mu < m$.  Large self-couplings in the gravitational sector, instead of a large number of new particles, might cause the running depicted in Fig.~\ref{fig:test}. That is, there might exist boundary values of the $c_i (\mu)$ at scale $\mu = \mu_*$ that lead to the observed large value of $c_1=M_P^{\,2}/16\pi$ at low energies. This would certainly require some anomalously large coefficients $c_i$, but current bounds are very weak and apply only at very low energies $\mu$. The strongest bounds come from experiments probing modifications of Newton's potential on distances of $\sim 0.1\,{\mbox{mm}}$ \cite{Stelle,Eoetwash}. One obtains $c_2(\mu \sim 10^{-3}{\mbox{eV}}) < 10^{61}$, with a similarly weak constraint holding for the coefficient of the other allowed four-derivative term $R^{\mu\nu}R_{\mu\nu}$ \cite{discussion}.

The phenomenology of the large $N$ model is described below. The most striking aspect of the model is that gravity is strong at a few TeV. In particular we expect that four dimensional black holes will be produced in high energy collisions of sufficient energy \cite{Eardley:2002re}. If these black holes are semi-classical they will decay via Hawking radiation, presumably primarily to the $N$ new degrees of freedom which overwhelmingly dominate the thermal phase space. Dvali {\it et al.} \cite{Dvali:2007wp} have emphasized that black holes formed from standard model particles might not decay into the $N$ new degrees of freedom. This would certainly be the case if the new particles all carry conserved charges. However, that would require of order $N$ additional gauge symmetries, which seems unattractive. Note, though, that decays of black holes of the smallest possible mass $M_{bh} \sim \mu_*$ (``quantum'' black holes) are not necessarily well described by semi-classical Hawking radiation. Quantum black holes might decay visibly, perhaps even to a small number of standard model particles \cite{Meade:2007sz}.

Experiments which detect showers caused by earth skimming neutrinos in the earth crust  \cite{Anchordoqui:2001cg} could still provide evidence for black holes that decay invisibly. If gravity is strong around 1 TeV, the probability for a high energy cosmic ray neutrino to collide with a nucleon and create a black hole is large. Earth skimming neutrinos within the standard model have a certain probability to convert to a lepton which escapes the crust of the earth and creates an observable shower. In scenarios of TeV gravity, some of these neutrinos will hit a nucleon and create a black hole which decays invisibly, reducing the earth skimming neutrino shower rate. The limit obtained in \cite{Anchordoqui:2001cg} (see also \cite{Anchordoqui:2003jr,Ringwald:2001vk,Kowalski:2002gb}) implies a bound on the cross-section  
\begin{eqnarray}
\sigma(\nu N \to qBH + X) ~<~ 0.5~{\rm TeV}^{-2}~.
\end{eqnarray}
Assuming that the parton level cross-section for quantum black holes is $\sigma = \mu_*^{-2}$, we get a bound $\mu_* > 1\, \mbox{TeV}$, which should really be considered to be an order of magnitude estimate. For $\mu_*$ of this size, quantum black hole production at the CERN LHC could have a cross section as large as
\begin{eqnarray}
 \sigma(pp  \to  qBH+ X) ~\sim~ 1 \times 10^5 \mbox{fb}~,
\end{eqnarray}
and will thus dominate the cross sections expected from the standard model. To the extent that small black holes behave as extremely hot, thermal objects, they will decay invisibly into the $10^{32}$ new degrees of freedom (barring an equal number of new conserved charges). However, quantum black holes might also decay visibly to a few standard model particles. In fact, the most common production process at LHC (e.g., gluon gluon $\to$ black hole) would in most cases leave the black hole with a net color charge. Confinement, or color neutrality, does not apply over length scales of order ${\rm TeV}^{-1}$, relevant for production and decay of quantum black holes. If the quantum black hole decays to a small number of particles, at least one of these particles will carry color and lead to a very energetic jet, which is potentially observable. A typical signature would be one high-$p_T$ jet plus missing energy. Besides colored black holes, small black holes with an electric charge will be produced frequently at the LHC. These charged black holes will decay most likely to one or two charged particles as well as a particle from the hidden sector. The charged particles are likely to be hadrons and would lead to one or two high-$p_T$ jets, but they could also be leptons. 

Depending on the parameters of the model, some semi-classical black holes could be produced at the LHC. The cross-section at the parton level is given by \cite{Eardley:2002re}:
\begin{eqnarray}
 \sigma (ij  \to  BH) ~=~ 4 \pi \frac{M^2_{bh}}{\mu_*^4}~,
\end{eqnarray}
where $M_{bh}$ is the black hole mass. If we take the scale of quantum gravity to be around $\sim 1$ TeV, this cross-section can be sizable for a semi-classical black hole mass of $\sim 3$ TeV. Taking into account that not all of the energy of the partons can be used in the formation of the black hole (see, e.g., \cite{Anchordoqui:2003ug}), the cross-section at the LHC is  $\sigma (pp  \to  BH+ X) \sim 2000 \ \mbox{fb}$ which for a luminosity of $100\  \mbox{fb}^{-1}$ would yield $2 \times 10^5$ semi-classical black holes.  As mentioned, these black holes will decay mostly invisibly into the new $N$ degrees of freedom unless there are a large number of new conserved charges. However, since it is likely that the black hole has a net color charge or an electric charge (as discussed in the previous paragraph), there will be at least one jet or a lepton in the final state, along with missing energy.

We have shown that the running of the gravitational coupling constant can be radically affected by a hidden sector with a large number of particles.  This implies that the scale for quantum gravity could be much different than the one obtained from naive dimensional analysis, i.e., $10^{19}\,$GeV. We discussed a specific model in which the scale of quantum gravity is in the TeV region. This model offers a solution to the hierarchy problem of the standard model and could lead to the production of quantum and semi-classical black holes at the LHC, with interesting signatures such as hard jet plus missing energy. It might also be testable through a deficit of earth skimming showers in high energy cosmic ray experiments such as AGASA.

\bigskip

\emph{Acknowledgments ---} We thank Gia Dvali for useful comments. This work is supported by the
Department of Energy under DE-FG02-96ER40969. XC is also supported by the Belgian Federal Office for Scientific, Technical and Cultural Affairs through the Interuniversity Attraction Pole P6/11.

\bigskip
\bigskip
%%%%%%%% End comments

%\newpage

%%%%%%%%%%%%%%%%%%%%%%%%%%%%%%%%%%%%%%%%%%%%%%%%%%%%%%%%%%%%%%%%%
%%%
%%%                     BIBLIOGRAPHY
%%%
%%%%%%%%%%%%%%%%%%%%%%%%%%%%%%%%%%%%%%%%%%%%%%%%%%%%%%%%%%%%%%%%%

\bigskip

%\newpage
%\vskip .75 in

\end{document}